\documentstyle[aps,twocolumn]{revtex}

\begin{document}
\draft

\title{A General Theory of Phase-Space Quasiprobability 
Distributions}
\author{C. Brif \cite{email1} \ and \ A. Mann \cite{email2}}
\address{Department of Physics, Technion -- Israel Institute of 
Technology, Haifa 32000, Israel}
\maketitle

\begin{abstract}
We present a general theory of quasiprobability distributions
on phase spaces of quantum systems whose dynamical symmetry
groups are (finite-dimensional) Lie groups. The family
of distributions on a phase space is postulated to satisfy
the Stratonovich-Weyl correspondence with a generalized
traciality condition. The corresponding family of the
Stratonovich-Weyl kernels is constructed explicitly. 
In the presented theory we use the concept of the generalized 
coherent states, that brings physical insight into the 
mathematical formalism. 
\end{abstract}

\pacs{03.65.Ca, 03.65.Fd, 03.65.Bz}

Since the introduction of the Wigner function in 1932 \cite{Wig32}, 
it has found numerous physical applications. Perhaps the most 
important is the phase-space formulation of quantum mechanics, that 
has its origins in the early work of Moyal \cite{Moy49}. 
In this formulation, a function on the phase space is associated with 
an operator on the Hilbert space, opening the way to formally 
representing quantum mechanics as a statistical theory on classical 
phase space. Various aspects of the phase-space formalism for basic 
quantum systems have been developed by a number of authors (e.g., 
Refs.\ \cite{Stra56,CaGl69,AgWo70,Bay78,Agar81,DAS94,GBVa,VaGB89,Kim}). 
More extensive lists of the literature on the subject can be found 
in review papers \cite{CaZa83,Hill84,Gade95}.

Besides the Wigner function $W$, other phase-space functions have 
been considered in the literature. In particular, the Husimi 
$Q$ function and the Glauber-Sudarshan $P$ function have found 
extensive applications in quantum optics.
Cahill and Glauber \cite{CaGl69} have shown that there exists
a whole family of phase-space functions parametrized by a number 
$s$; the values $+1$, $0$, and $-1$ of $s$ correspond to the $Q$, 
$W$, and $P$ functions, respectively. These phase-space functions
are known as quasiprobability distributions (QPDs), as they play 
in quantum mechanics a role similar to that of genuine probability 
distributions in classical statistical mechanics.

The phase-space formalism has been applied successfully to the
description of a spinless quantum particle and a mode of the 
quantized radiation field (modelled by a quantum harmonic 
oscillator). The corresponding phase space is ${\sf R}^2$ (or,
equivalently, the complex plane). A generalization of this 
description to a set of $N$ independent particles or harmonic 
oscillators in a $p$-dimensional world is straightforward 
\cite{Gade95}.
A more complicated problem is the phase-space description
of spin. A number of authors have used different approaches 
to the construction of the Wigner function for spin 
\cite{Agar81,DAS94,VaGB89,KaSu69,Fron79,MoON83,Gilm87,%
Woot87,Leonh}. 
The explicit expressions for the $Q$, $W$, and $P$ functions 
for arbitrary spin were first obtained by Agarwal 
\cite{Agar81}, who used the spin coherent-state 
representation \cite{ACGT72,Per} and the Fano multipole 
operators \cite{Fano}. 
V\'{a}rilly and Gracia-Bond\'{\i}a \cite{VaGB89} have shown that 
the spin coherent-state approach is equivalent to a general
mathematical formalism based on the Stratonovich-Weyl (SW)
correspondence \cite{Stra56} and on the concept of twisted 
product \cite{Bay78}.

In the present Letter we develop a general theory of QPDs
on phase spaces of quantum systems whose dynamical symmetry
groups are (finite-dimensional) Lie groups. This theory
can be viewed as a generalization of the Cahill-Glauber 
QPDs (related to the Heisenberg-Weyl group) to arbitrary Lie 
groups. 
We make clear that the structure of the family of the 
phase-space QPDs for a Lie group is determined by the 
group covariance and the traciality conditions.

Let $G$ be a Lie group (connected and simply connected, with 
finite dimension $n$), that is the dynamical symmetry group 
of a given quantum system. Let $T$ be a unitary irreducible 
representation of $G$ acting on the Hilbert space ${\cal H}$. 
By choosing a fixed normalized reference state 
$|\psi_0\rangle \in {\cal H}$, one can define the system of 
coherent states $\{|\psi_g\rangle\}$:
\begin{equation}
|\psi_g\rangle = T(g) |\psi_0\rangle , \hspace{0.8cm}  g \in G .
\end{equation}
The isotropy subgroup $H \subset G$ consists 
of all the group elements $h$ that leave the reference state 
invariant up to a phase factor,
\begin{equation}
T(h) |\psi_{0}\rangle = e^{i\phi(h)} |\psi_{0}\rangle , 
\hspace{0.8cm} | e^{i\phi(h)} | =1 ,
\hspace{0.3cm} h \in H .                
\end{equation}
For every element $g \in G$, there is a unique decomposition 
of $g$ into a product of two group elements, one in $H$ and 
the other in the coset space $X = G/H$,
\begin{equation}
g = \Omega h , \hspace{0.8cm} g \in G , \;\; h \in H , \;\; 
\Omega \in X .  
\end{equation}
It is clear that group elements $g$ and $g'$ with different $h$ 
and $h'$ but with the same $\Omega$ produce coherent states 
which differ only by a phase factor: 
$|\psi_{g}\rangle = e^{i\delta} |\psi_{g'}\rangle$, where 
$\delta =\phi(h) -\phi(h')$. 
Therefore a coherent state $|\Omega\rangle \equiv
|\psi_{\Omega}\rangle$ is 
determined by a point $\Omega = \Omega(g)$ in the coset 
space $G/H$. As a direct consequence of Schur's lemma, one
obtains the identity resolution in terms of the coherent 
states:
\begin{equation}
\int_{X} d \mu(\Omega) |\Omega\rangle\langle\Omega| = I ,
\end{equation}
where $d \mu(\Omega)$ is the invariant integration measure
on $X = G/H$, the integration is over the whole manifold $X$, 
and $I$ is the identity operator on ${\cal H}$.

An important class of coherent-state systems 
corresponds to the coset spaces $X = G/H$ which are homogeneous
K\"{a}hlerian manifolds. Then $X$ can be considered as the 
phase space of a classical dynamical system, and the mapping 
$\Omega \rightarrow |\Omega\rangle\langle\Omega|$ 
is the quantization for this system \cite{Berez}. The standard 
(or maximal-symmetry) systems of the coherent states correspond to 
the cases when an extreme state of the representation Hilbert 
space (e.g., the vacuum state of an oscillator or the lowest
spin state) is chosen as the reference state. In general, this 
choice of the reference state leads to systems consisting of 
states with properties ``closest to those of classical states'' 
\cite{Per}. In what follows we will consider the coherent 
states of maximal symmetry and assume that the phase 
space of the quantum system is a homogeneous K\"{a}hlerian 
manifold $X = G/H$, each point of which corresponds to a 
coherent state $|\Omega\rangle$. In particular, the Glauber
coherent states of the Heisenberg-Weyl group H$_{3}$ are 
defined on the complex plane 
${\sf R}^2 = {\rm H}_3 / {\rm U}(1)$, and the spin coherent
states are defined on the unit sphere 
${\sf S}^2 = {\rm SU}(2) / {\rm U}(1)$.
In a more rigorous mathematical language of Kirillov's theory
\cite{Kiril}, the phase space $X$ is the coadjoint orbit
associated with the unitary irreducible representation $T$
of the group $G$ on the Hilbert space ${\cal H}$.

The idea of the phase-space formalism is as follows. Let $A$
be an operator on ${\cal H}$. Then $A$ can be mapped by
a family of functions (quasiprobability distributions) 
$F_{A}^{(s)}(\Omega)$ onto the phase space $X$. (The
index $s$ that labels functions in the family will be
determined shortly).
The function $F_{A}^{(s)}(\Omega)$ is called the SW image of 
$A$, if it satisfies the properties 
known as the SW correspondence \cite{Stra56}:
\begin{mathletters}
\begin{enumerate}
\item[(0)] Linearity: $A \rightarrow F_{A}^{(s)}(\Omega)$ 
is one-to-one linear map.
\item[(i)] Reality: 
\begin{equation}
F_{A^{\dagger}}^{(s)}(\Omega) = [F_{A}^{(s)}(\Omega)]^{\ast} .
\label{real} 
\end{equation}
\item[(ii)] Standardization: 
\begin{equation}
\int_{X} d \mu(\Omega) F_{A}^{(s)}(\Omega) = {\rm Tr}\, A .
\label{stand}
\end{equation}
\item[(iii)] Covariance: 
\begin{equation}
F_{g \cdot A}^{(s)}(\Omega) = F_{A}^{(s)}(g^{-1} \Omega) , 
\label{covar}
\end{equation}
where $g \cdot A \equiv T(g) A T(g)^{-1}$.
\item[(iv)] Traciality:
\begin{equation}
\int_{X} d \mu(\Omega) F_{A}^{(s)}(\Omega) 
F_{B}^{(-s)}(\Omega) = {\rm Tr}\, (AB) .
\label{trac}
\end{equation}
\end{enumerate}
\end{mathletters}
These conditions have a clear physical meaning.
The linearity and the traciality conditions are related to the
statistical interpretation of the theory. If $B$ is the density 
matrix, then the traciality condition (\ref{trac}) assures that
the statistical average of the phase-space distribution
$F_{A}$ coincides with the quantum expectation value of the
operator $A$. Equation (\ref{trac}) is actually a generalization
of the usual traciality condition \cite{Stra56,VaGB89}, as it
holds for any $s$ and not only for the Wigner case $s=0$.
The reality condition (\ref{real}) means that
if $A$ is self-adjoint, then $F_{A}^{(s)}(\Omega)$ is real.
The condition (\ref{stand}) is a natural normalization,
which means that the image of the identity operator $I$ is
the constant function $1$. The covariance condition
(\ref{covar}) means that the phase-space formulation
must explicitly express the symmetry of the system.

The linearity is taken into account, if we implement the map
$A \rightarrow F_{A}^{(s)}(\Omega)$ by the generalized Weyl 
rule
\begin{equation}
F_{A}^{(s)}(\Omega) = {\rm Tr}\, [A \Delta^{(s)}(\Omega)] ,
\label{gwr}
\end{equation}
where $\{\Delta^{(s)}(\Omega)\}$ is a family (labelled by $s$)
of operator-valued functions on the phase space $X$. 
These operators are referred to as the SW kernels. 
The generalized traciality condition (\ref{trac}) is taken 
into account, if we define the inverse of the generalized 
Weyl rule (\ref{gwr}) as
\begin{equation}
A = \int_{X} d \mu(\Omega) F_{A}^{(s)}(\Omega) 
\Delta^{(-s)}(\Omega) . \label{invmap}
\end{equation}
Now, the conditions (\ref{real})-(\ref{covar}) of the SW 
correspondence for $F_{A}^{(s)}(\Omega)$ can be translated 
into the following conditions on the SW kernel 
$\Delta^{(s)}(\Omega)$:
\begin{mathletters}
\begin{eqnarray}
\rm{(i)} & & \;\;\; \Delta^{(s)}(\Omega) = 
[\Delta^{(s)}(\Omega)]^{\dagger}
\;\;\;\;\;\; \forall \Omega \in X .
\label{real1} \\
\rm{(ii)} & & \;\;\; \int_{X} d \mu(\Omega) \Delta^{(s)}(\Omega) 
= I . \label{stand1} \\
\rm{(iii)} & & \;\;\; \Delta^{(s)}(g\Omega) = 
T(g) \Delta^{(s)}(\Omega) T(g)^{-1} . 
\label{covar1}
\end{eqnarray}
\end{mathletters}

Substituting the inverted maps (\ref{invmap}) for $A$ and 
$B$ into the generalized traciality condition (\ref{trac}),
we obtain the relation between the QPDs with different values
of the index $s$: 
\begin{eqnarray}
& & F_{A}^{(s)}(\Omega) = \int_{X} d \mu(\Omega') 
K_{s,s'}(\Omega,\Omega') F_{A}^{(s')}(\Omega') , 
\label{genrel} \\
& & K_{s,s'}(\Omega,\Omega') \equiv 
{\rm Tr}\, [ \Delta^{(s)}(\Omega) \Delta^{(-s')}(\Omega')] .
\label{Kssfun}
\end{eqnarray}
If we take in equation (\ref{genrel}) $s=s'$ and take into account
the arbitrariness of $A$, we obtain the following relation
\begin{equation}
\Delta^{(s)}(\Omega) = \int_{X} d \mu(\Omega') K(\Omega,\Omega')
\Delta^{(s)}(\Omega') ,  \label{specrel}
\end{equation}
where the function
\begin{equation}
K(\Omega,\Omega') = {\rm Tr}\, [ \Delta^{(s)}(\Omega) 
\Delta^{(-s)}(\Omega')]
\label{Kfun}
\end{equation}
behaves like the delta function on the manifold $X$.

Now, our problem is to find the explicit form of the 
SW kernel $\Delta^{(s)}(\Omega)$ that satisfies the 
conditions (\ref{real1})-(\ref{covar1}) and (\ref{specrel}).
We start by considering the Hilbert space $L^{2}(X,\mu)$ of 
square-integrable functions $u(\Omega)$ on $X$ with the 
invariant measure $ d \mu$. The representation $T$ of the 
Lie group $G$ on $L^{2}(X,\mu)$ is defined as 
\begin{equation}
T(g) u(\Omega) = u(g^{-1}\Omega) . 
\end{equation}
The eigenfunctions $Y_{\nu}(\Omega)$ of the Laplace-Beltrami
operator \cite{BaRa86} form a complete orthonormal basis in
$L^{2}(X,\mu)$:
\begin{mathletters}
\begin{eqnarray}
& & \sum_{\nu} Y_{\nu}^{\ast}(\Omega) Y_{\nu}(\Omega') 
= \delta(\Omega-\Omega') , \\
& & \int_{X} d \mu(\Omega) Y_{\nu}^{\ast}(\Omega) 
Y_{\nu'}(\Omega) = \delta_{\nu \nu'} .
\end{eqnarray}
\end{mathletters}
The functions $Y_{\nu}(\Omega)$ are called the harmonic 
functions \cite{comment1}, and $\delta(\Omega-\Omega')$ 
is the delta function in $X$ with respect to the measure
$ d \mu$. The eigenfunctions $Y_{\nu}(\Omega)$ are linear
combinations of matrix elements $T_{\nu \nu'}(g)$. 
Therefore, the transformation rule for the harmonic 
functions is \cite{BaRa86}
\begin{equation}
T(g) Y_{\nu}(\Omega) = Y_{\nu}(g^{-1}\Omega)
= \sum_{\nu'} T_{\nu' \nu}(g) Y_{\nu'}(\Omega) .
\label{transrule}
\end{equation}
The function $|\langle \Omega | \Omega' \rangle|^2$ is 
symmetric in $\Omega$ and $\Omega'$. Therefore, its 
expansion in the orthonormal basis must be of the form
\begin{equation}
|\langle \Omega | \Omega' \rangle|^2 = \sum_{\nu} \tau_{\nu} 
Y_{\nu}^{\ast}(\Omega) Y_{\nu}(\Omega') = \sum_{\nu}
\tau_{\nu} Y_{\nu}^{\ast}(\Omega') Y_{\nu}(\Omega) ,
\label{ooexpansion}
\end{equation}
where $\tau_{\nu}$ are real positive coefficients.
Since $|\langle \Omega | \Omega' \rangle|^2$ is real and
$Y_{\nu}^{\ast}(\Omega) = e^{ i \phi(\nu)} 
Y_{\tilde{\nu}}(\Omega)$, the
coefficients $\tau_{\nu}$ must be invariant under this 
index transformation: $\tau_{\nu} = \tau_{\tilde{\nu}}$.
Since $\langle \Omega | \Omega' \rangle = 
\langle g\Omega | g\Omega' \rangle$, the coefficients 
$\tau_{\nu}$ must be invariant under the index 
transformation of Eq.\ (\ref{transrule}): 
$\tau_{\nu} = \tau_{\nu'}$.

Let us now define the set of operators $\{ D_{\nu} \}$ 
on ${\cal H}$:
\begin{equation}
D_{\nu} \equiv \omega_{\nu} \int_{X} d \mu(\Omega) 
Y_{\nu}(\Omega) | \Omega \rangle \langle \Omega | ,
\end{equation}
where $\omega_{\nu}$ are real coefficients to be 
determined from the normalization condition. 
Using the expression (\ref{ooexpansion}), we obtain the
orthogonality condition
\begin{equation}
{\rm Tr}\, (D_{\nu} D_{\nu'}^{\dagger}) = 
( \tau_{\nu} \omega_{\nu}^2 )\, \delta_{\nu \nu'} .
\end{equation}
The proper normalization is then obtained by taking
\begin{equation}
\omega_{\nu}^2 = 1/\tau_{\nu} .
\end{equation}
Note that $\omega_{\nu}$ is defined only up to a sign,
$\omega_{\nu} = \pm \tau_{\nu}^{-1/2}$.
Using (\ref{ooexpansion}), we also obtain the relation
\begin{equation}
\omega_{\nu} \langle \Omega | D_{\nu} | \Omega \rangle =
Y_{\nu}(\Omega) .
\end{equation}
The coefficients $\omega_{\nu}$ satisfy the same 
invariance conditions as $\tau_{\nu}$ (up to a choice
of the sign). Therefore, $D_{\nu}$ are the tensor 
operators whose transformation rule is the same as 
for the harmonic functions $Y_{\nu}(\Omega)$:
\begin{equation}
T(g) D_{\nu} T(g)^{-1} = 
\sum_{\nu'} T_{\nu' \nu}(g) D_{\nu'} .
\label{dtrule}
\end{equation}

Now we are able to find the SW kernel $\Delta^{(s)}(\Omega)$
with all the desired properties. Specifically, let us define
\cite{comment2}
\begin{equation}
\Delta^{(s)}(\Omega) \equiv \sum_{\nu} f(s;\tau_{\nu})
Y_{\nu}^{\ast}(\Omega) D_{\nu} .
\label{kerneldef}
\end{equation}
Here $f(s;\tau_{\nu})$ is a function of $\tau_{\nu}$ and
of the index $s$. 
We assume that $f$ possesses the invariance properties of 
$\tau_{\nu}$. The reality condition (\ref{real1})
is then satisfied if $f(s;\tau_{\nu})$ is a real-valued 
function. Therefore, we can consider only real values of the 
index $s$. Then it is sufficient to use the convention in 
which $s \in [-1,1]$. 
Next we consider the standardization condition (\ref{stand1}).
It can be verified that there exists some $\nu_{0}$ such that
\cite{comment3}
\begin{equation}
\int_{X} d \mu(\Omega) Y_{\nu}(\Omega) \propto 
\delta_{\nu \nu_{0}} .
\label{delta0}
\end{equation}
We also are able to show that $\tau_{\nu_{0}} = 1$.
Then the standardization condition (\ref{stand1}) is
satisfied if 
\begin{equation}
f(s;1) = \omega_{\nu_{0}} = \pm 1 , \;\;\;\; \forall s .
\label{stand2}
\end{equation}
The covariance condition (\ref{covar1}) is guaranteed by
virtue of the transformation rules (\ref{transrule})
and (\ref{dtrule}) and by the invariance of $\tau_{\nu}$
under these index transformations.

In order to satisfy the relation (\ref{specrel}), the
function $K(\Omega,\Omega')$ of Eq.\ (\ref{Kfun}) must
be the delta function in $X$ with respect to the measure
$ d \mu$,
\begin{equation}
K(\Omega,\Omega') = 
\sum_{\nu} Y_{\nu}^{\ast}(\Omega) Y_{\nu}(\Omega') 
= \delta(\Omega-\Omega') .
\end{equation}
This result is valid if
\begin{equation}
f(s;\tau_{\nu}) f(-s;\tau_{\nu}) = 1 .
\end{equation}
This property is satisfied only by the exponential
function, i.e.,
\begin{equation}
f(s;\tau_{\nu}) = \pm [f(\tau_{\nu})]^{s} . \label{explaw}
\end{equation}
Note that the standardization condition (\ref{stand2})
then reads $f(1) = 1$. The double-valuedness of type 
(\ref{explaw}) was pointed out by V\'{a}rilly and 
Gracia-Bond\'{\i}a \cite{VaGB89} who considered the
Wigner function for spin. The exact form of the 
function $f(\tau_{\nu})$ can be determined if we 
define for $s=1$
\begin{equation}
\Delta^{(1)}(\Omega) \equiv | \Omega \rangle \langle \Omega | .
\end{equation}
Then we obtain $\pm f(\tau_{\nu}) = 1/\omega_{\nu} =
\pm \tau_{\nu}^{1/2}$, i.e.,
\begin{equation}
f(\tau_{\nu}) = \sqrt{\tau_{\nu}} .
\end{equation}
Obviously, the standardization condition
$f(1)=1$ is satisfied.
This result concludes the construction of the generalized
SW kernel. It is evident that the properties of the kernels
are completely determined by the harmonic functions on the
corresponding manifold and by the coherent states that form
this manifold. 
We also note that the function 
$K_{s,s'}(\Omega,\Omega')$ of Eq.\ (\ref{Kssfun}) is
given by
\begin{equation}
K_{s,s'}(\Omega,\Omega') = \sum_{\nu} \tau_{\nu}^{(s-s')/2}
Y_{\nu}^{\ast}(\Omega) Y_{\nu}(\Omega') ,
\end{equation}
and it clearly satisfies the condition (\ref{genrel}).

In order to avoid the double-valuedness of type (\ref{explaw}), 
we adopt the convention with sign ``+'', i.e., 
$\omega_{\nu} = + \tau_{\nu}^{-1/2}$. Then we can write
the generalized QPDs on the phase space as
\begin{eqnarray}
& & F_{A}^{(s)}(\Omega) = \sum_{\nu} \tau_{\nu}^{s/2} 
{\cal A}_{\nu} Y_{\nu}(\Omega) ,  \\
& & {\cal A}_{\nu} \equiv {\rm Tr}\, (A D_{\nu}^{\dagger})
= \tau_{\nu}^{-1/2} \int_{X} d \mu(\Omega) 
Y_{\nu}^{\ast}(\Omega) \langle \Omega |A| \Omega \rangle . 
\end{eqnarray}
In particular, for $s=1$, we obtain the $Q$ function
(Berezin's covariant symbol \cite{Berez}):
\begin{equation}
Q_{A}(\Omega) \equiv F_{A}^{(1)}(\Omega) = 
\langle \Omega |A| \Omega \rangle .
\end{equation}
For $s=-1$, we obtain the $P$ function
(Berezin's contravariant symbol \cite{Berez}):
\begin{eqnarray}
& & P_{A}(\Omega) \equiv F_{A}^{(-1)}(\Omega) = 
\sum_{\nu} \omega_{\nu} {\cal A}_{\nu} Y_{\nu}(\Omega) , \\
& & A = \int_{X} d \mu(\Omega) P_{A}(\Omega) 
| \Omega \rangle \langle \Omega | .
\end{eqnarray}
The functions $P$ and $Q$ are counterparts in the
traciality condition (\ref{trac}).
Perhaps the most important QPD corresponds to $s=0$,
because this function is ``self-conjugate'' in the sense
that it is the counterpart of itself in the traciality
condition (\ref{trac}). It is natural to call the
QPD with $s=0$ the generalized Wigner function:
\begin{equation}
W_{A}(\Omega) \equiv F_{A}^{(0)}(\Omega) = 
\sum_{\nu} {\cal A}_{\nu} Y_{\nu}(\Omega) .
\end{equation}

In conclusion, we have developed the general 
group-theoretical formalism of the phase-space 
QPDs. More details and examples of the QPDs on
phase spaces of physical systems will be
presented elsewhere.

This work was supported by the Fund for Promotion
of Research at the Technion and by the Technion
VPR Fund --- The R. and M. Rochlin Research Fund.


\begin{references}
%
\bibitem[*]{email1} E-mail: costya@physics.technion.ac.il
\bibitem[\dag]{email2} E-mail: ady@physics.technion.ac.il
%
\bibitem{Wig32} E. Wigner, Phys. Rev. {\bf 40}, 749 (1932).
%
\bibitem{Moy49} J. E. Moyal, Proc. Cambridge Philos. Soc. 
{\bf 45}, 99 (1949).
%
\bibitem{Stra56} R. L. Stratonovich, Zh. Eksp. Teor. Fiz. 
{\bf 31}, 1012 (1956) 
[Sov. Phys. JETP {\bf 4}, 891 (1957)].
%
\bibitem{CaGl69} K. E. Cahill and R. J. Glauber, Phys. Rev. 
{\bf 177}, 1857 (1969); {\bf 177}, 1882 (1969).
%
\bibitem{AgWo70} G. S. Agarwal and E. Wolf, Phys. Rev. D
{\bf 2}, 2161 (1970); {\bf 2}, 2187 (1970); {\bf 2}, 2206 (1970).
%
\bibitem{Bay78} F. Bayen, M. Flato, C. Fronsdal, A. Lichnerowicz,
and D. Sternheimer, Ann Phys. (N.Y.) {\bf 111}, 61 (1978);
{\bf 111}, 111 (1978).
%
\bibitem{Agar81} G. S. Agarwal, Phys. Rev. A {\bf 24}, 2889 (1981).
\bibitem{DAS94} J. P. Dowling, G. S. Agarwal, and W. P. Schleich,
Phys. Rev. A {\bf 49}, 4101 (1994).
%
\bibitem{GBVa} J. M. Gracia-Bond\'{\i}a, Phys. Rev. A {\bf 30}, 691 
(1984); J. M. Gracia-Bond\'{\i}a and J. C. V\'{a}rilly, J. Math. Phys.
{\bf 29}, 869 (1988); {\bf 29}, 880 (1988).
%
\bibitem{VaGB89} J. C. V\'{a}rilly and J. M. Gracia-Bond\'{\i}a,
Ann Phys. (N.Y.) {\bf 190}, 107 (1989).
%
\bibitem{Kim} Y. S. Kim and M. E. Noz, {\em Phase Space Picture
of Quantum Mechanics} (World Scientific, Singapore, 1991).
%
\bibitem{CaZa83} P. Carruthers and F. Zachariasen, Rev. Mod. Phys.
{\bf 55}, 245 (1983).
%
\bibitem{Hill84} M. Hillery, R. F. O'Connell, M. O. Scully, and
E. P. Wigner, Phys. Rep. {\bf 106}, 121 (1984).
%
\bibitem{Gade95} M. Gadella, Fortschr. Phys. {\bf 43}, 229 (1995).
%
\bibitem{KaSu69} D. M. Kaplan and G. C. Summerfield, Phys. Rev.
{\bf 187}, 639 (1969).
%
\bibitem{Fron79} C. Fronsdal, Rep. Math. Phys. {\bf 15}, 111 (1979).
%
\bibitem{MoON83} C. Moreno and P. Ortega-Navarro, Lett. Math. Phys.
{\bf 7}, 181 (1983).
%
\bibitem{Gilm87} R. Gilmore, in {\em Lecture Notes in Physics}, 
Vol. {\bf 278}, edited by Y. S. Kim and W. W. Zachary (Springer,
Berlin, 1987), p. 211.
%
\bibitem{Woot87} W. K. Wootters, Ann. Phys. (N.Y.) {\bf 176},
1 (1987).
\bibitem{Leonh} U. Leonhardt, Phys. Rev. Lett. {\bf 74},
4101 (1995); Phys. Rev. A {\bf 53}, 2998 (1996).
%
\bibitem{ACGT72} F. T. Arecchi, E. Courtens, R. Gilmore, 
and H. Thomas, Phys. Rev. A {\bf 6}, 2211 (1972).
%
\bibitem{Per} A. M. Perelomov, Commun. Math. Phys. {\bf 26}, \
222 (1972); {\em Generalized Coherent States and Their 
Applications} (Springer, Berlin, 1986).
%
\bibitem{Fano} U. Fano, Phys. Rev. {\bf 90}, 577 (1953).
%
\bibitem{Berez} F. A. Berezin, Commun. Math. Phys. {\bf 40}, 
153 (1975).
%
\bibitem{Kiril} A. A. Kirillov, {\em Elements of the Theory of 
Representations} (Springer, Berlin, 1976).
%
\bibitem{BaRa86} A. O. Barut and R. Raczka, {\em Theory of 
Group Representations and Applications}, 2nd ed. 
(World Scientific, Singapore, 1986), Chap. 15.
%
\bibitem{comment1} The index $\nu$ is multiple; it has one 
discrete part, while the other part is discrete for compact 
manifolds and continuous for noncompact manifolds. 
In the last case the summation over $\nu$ includes an 
integration with an appropriate measure and the symbol
$\delta_{\nu \nu'}$ includes some Dirac delta function.
For more details see Ref.\ \cite{BaRa86}. 
For conciseness, we omit these details in our formulas.
%
\bibitem{comment2} One can see that the SW kernel 
(\ref{kerneldef}) is a generalization of the Cahill-Glauber
kernel for a harmonic oscillator \cite{CaGl69,AgWo70} and
the Agarwal kernel for spin \cite{Agar81}. We show that the
construction of the generalized kernel (\ref{kerneldef}) is 
adjusted to satisfy the SW correspondence.   
%
\bibitem{comment3} As has been mentioned before 
\cite{comment1}, for noncompact manifolds the symbol
$\delta_{\nu \nu'}$ actually includes some Dirac 
delta function.


\end{references}
\end{document}